\documentclass[aps,prb,twocolumn,amsmath,amssymb,groupedaddress,uplatex]{revtex4-2}

\usepackage{graphicx}
\usepackage{dcolumn}
\usepackage{bm}
\usepackage{multirow}
\usepackage{braket}
\usepackage{color}
\usepackage{ulem}
\usepackage{booktabs}

\begin{document}
\title{
Mechanism of antisymmetric spin polarization in centrosymmetric multiple-$Q$ magnets \\
based on bilinear and biquadratic spin cross products
}
\author{Satoru~Hayami}
\affiliation{Department of Applied Physics, the University of Tokyo, Tokyo 113-8656, Japan}

\begin{abstract}
We investigate how to engineer an antisymmetric spin-split band structure under spin density waves with finite ordering wave vectors in centrosymmetric systems without the relativistic spin-orbit coupling. 
On the basis of a perturbative analysis for the spin-charge coupled model in centrosymmetric itinerant magnets, we show that nonzero chiral-type bilinear and biquadratic spin cross products in momentum space under the magnetic orderings are related to an antisymmetric spin polarization in the electronic band structure. 
We apply the derived formula to the single-$Q$ cycloidal spiral and double-$Q$ noncoplanar states including the meron-antimeron and skyrmion crystals. 
Our results present a clue to realize a giant antisymmetric spin splitting driven by magnetic phase transitions in the centrosymmetric lattice structures without the spin-orbit coupling. 
\end{abstract}

\maketitle

\section{Introduction}

Symmetry is an important factor to determine physical properties of solids. 
Among them, spatial inversion symmetry has drawn considerable interest in condensed matter physics, since its breaking gives rise to fascinating physical phenomena, such as a spontaneous electric polarization and nonreciprocal transport~\cite{Aizu_RevModPhys.34.550,Resta_RevModPhys.66.899,van2008multiferroicity,ideue2017bulk,tokura2018nonreciprocal}. 
The breaking of spatial inversion symmetry also leads to an antisymmetric spin polarization in terms of the wave vectors in electronic band structures, which has been often found in the noncentrosymmetric crystals with the strong relativistic spin-orbit coupling~\cite{Dresselhaus_Dresselhaus_Jorio,frigeri2004spin,palazzese2018strong}, such as polar crystals with the Rashba-type spin-orbit coupling~\cite{rashba1960properties,ishizaka2011giant,Bahramy_PhysRevB.84.041202,sunko2017maximal}, chiral crystals with the Weyl-type spin-orbit coupling~\cite{Takashima_PhysRevB.94.235117,Wang_PhysRevA.97.011605}, and other noncentrosymmetric crystals with the Ising-type spin-orbit coupling~\cite{Zhu_PhysRevB.84.153402,wang2012electronics,ugeda2014giant,Andor_PhysRevX.4.011034}. 
The antisymmetric spin polarization becomes a source of spin-related parity-violating physical phenomena~\cite{Fu_PhysRevLett.115.026401,Kozii_PhysRevLett.115.207002,Venderbos_PhysRevB.94.180504,Hayami_PhysRevLett.122.147602}, such as the spin Hall effect~\cite{murakami2003dissipationless,Murakami_PhysRevLett.93.156804,Sinova_PhysRevLett.92.126603,Fujimoto_doi:10.1143/JPSJ.75.083704,fujimoto2007fermi} and the Edelstein effect~\cite{edelstein1990spin,Yip_PhysRevB.65.144508,Fujimoto_PhysRevB.72.024515,yoda2018orbital,Massarelli_PhysRevB.100.075136}. 

The above parity-violating phenomena also occur under the centrosymmetric crystal structures once the spatial inversion symmetry is broken by a spontaneous phase transition through the electron correlation~\cite{tokura2014multiferroics,Fu_PhysRevLett.115.026401,hayami2016emergent}. 
Especially, magnetic phase transitions to the noncollinear and noncoplanar magnetic ordered states actualize the antisymmetric spin-orbit interaction even without the relativistic spin-orbit coupling. 
One of the examples is the inverse Dzyaloshinskii-Moriya mechanism where the spin vector chirality in noncollinear magnets produces the electric polarization~\cite{Katsura_PhysRevLett.95.057205,Mostovoy_PhysRevLett.96.067601,SergienkoPhysRevB.73.094434,Harris_PhysRevB.73.184433,tokura2014multiferroics,cardias2020first}. 
Another example is the emergence of the electric polarization and the nonreciprocal transport owing to nonzero spin scalar chirality in noncoplanar magnets~\cite{Bulaevskii_PhysRevB.78.024402,batista2016frustration,hayami2020phase,ishizuka2020anomalous,hayami2021locking}. 
Besides, the origin of the antisymmetric spin polarization under the noncollinear and noncoplanar spin configurations has been microscopically studied based on augmented multipoles~\cite{Hayami_PhysRevB.101.220403,Hayami_PhysRevB.102.144441}, which is compatible with magnetic point group symmetry~\cite{Hayami_PhysRevB.98.165110,Watanabe_PhysRevB.98.245129,Yuan_PhysRevMaterials.5.014409,Yatsushiro_PhysRevB.104.054412}. 

Designing and engineering the antisymmetric spin-split band structure under the noncollinear and noncoplanar spin textures stimulate a further exploration of functional materials with a giant spin splitting even in the absence of the relativistic spin-orbit coupling in the centrosymmetric lattice structures.
It has an advantage of opening up the option of candidate materials so as to include light-element materials and 3$d$ transition metal oxides in addition to conventional heavy-element ones with the strong spin-orbit coupling. 
Such an extension of candidate materials will be useful for a future realization of high-efficient electronics and spintronics devices. 

From the energetic point of view, there are various mechanisms to stabilize noncollinear and noncoplanar magnetic orderings that break the spatial inversion symmetry, such as a spiral state and a skyrmion crystal, in the centrosymmetric lattice structures: the frustrated exchange interactions in insulating magnets~\cite{tamura2011first,Okubo_PhysRevLett.108.017206,leonov2015multiply,Lin_PhysRevB.93.064430,Hayami_PhysRevB.93.184413,Takashima_PhysRevB.94.134415,Lohani_PhysRevX.9.041063,amoroso2020spontaneous,Hayami_PhysRevB.103.224418} and the multiple-spin interactions and magnetic anisotropy in itinerant magnets~\cite{Ozawa_PhysRevLett.118.147205,Hayami_PhysRevB.95.224424,Hayami_PhysRevB.99.094420,Wang_PhysRevLett.124.207201,Hayami_PhysRevB.103.054422,hayami2021topological,yambe2021skyrmion,Hayami_10.1088/1367-2630/ac3683}. 
Considering that the skyrmion crystal and other various noncoplanar magnetic states are described by a superposition of the multiple-$Q$ spiral waves, one can expect a possibility of realizing the giant antisymmetric spin splitting without relying on the relativistic spin-orbit coupling in centrosymmetric multiple-$Q$ spiral magnets. 
However, the relationship between electronic band structures and multiple-$Q$ spiral spin textures has not been fully understood yet. 

In the present paper, we study a microscopic mechanism of the spin-dependent antisymmetric band modulation in the single-$Q$ and multiple-$Q$ spiral states to open another route of noncentrosymmetric spin-orbit-coupled physics in inversion-symmetric materials with negligibly small atomic spin-orbit coupling. 
We derive effective momentum-dependent chiral-type bilinear and biquadratic spin cross products in momentum space under the magnetic orderings by performing the perturbative expansion with respect to the exchange coupling in the classical Kondo lattice model. 
The derived expressions indicate that the antisymmetric spin polarization appears when the magnetic orderings with nonzero bilinear and biquadratic spin cross product occur. 
Moreover, the expressions provide necessary multiple-$Q$ spin modulations to cause the momentum-dependent antisymmetric spin polarization in the band structure. 
We test the expressions to the single-$Q$ state on a one-dimensional chain and the double-$Q$ states on a two-dimensional square lattice. 
We also apply the expressions to the skyrmion-hosting centrosymmetric magnet GdRu$_2$Si$_2$.
The examples include the square-shaped meron-antimeron and skyrmion crystals. 
The present results are ubiquitously applied to any magnetic textures with finite ordering wave vectors in any lattice systems, which will be useful to extend the scope of materials with a giant antisymmetric spin splitting in centrosymmetric magnets even without the spin-orbit coupling. 

This article is organized in the following way: 
In Sec.~\ref{sec:Model}, we introduce the classical Kondo lattice model as one of the fundamental models in itinerant magnets. 
In Sec.~\ref{sec:Effective chiral interactions}, we show a derivation of effective momentum-dependent chiral-type bilinear and biquadratic spin cross products by using the perturbation expansion for the Kondo lattice model in terms of the exchange coupling between itinerant electrons and localized spins. 
We discuss the antisymmetric spin-split band structure under the single-$Q$ and multiple-$Q$ spiral states on the basis of the derived formula in Sec.~\ref{sec:Antisymmetric spin splitting in single-$Q$ and multiple-$Q$ states}. 
We also discuss the relevant materials and physical phenomena in Sec.~\ref{sec:Discussion}.
Section~\ref{sec:Summary} is devoted to the summary.

\section{Model}
\label{sec:Model}

We study a spin-charge coupled system consisting of itinerant electrons and localized spins with itinerant magnets in mind. 
For that purpose, we adopt the classical Kondo lattice (double exchange) model with the exchange coupling between itinerant electron spins and localized spins, which is one of the underlying models to exhibit a plethora of multiple-$Q$ spiral states~\cite{Ozawa_doi:10.7566/JPSJ.85.103703,Ozawa_PhysRevLett.118.147205,Hayami_PhysRevB.99.094420,Mohanta_PhysRevB.100.064429,Wang_PhysRevLett.124.207201,Kathyat_PhysRevB.103.035111,hayami2020phase}. 
The Hamiltonian is given by 
\begin{align}
\label{eq:Ham_KLM}
\mathcal{H}= -\sum_{i, j,  \sigma} t_{ij} c^{\dagger}_{i\sigma}c_{j \sigma}
+J \sum_{i, \sigma, \sigma'} c^{\dagger}_{i\sigma} \bm{\sigma}_{\sigma \sigma'} c_{i \sigma'}
\cdot \bm{S}_i, 
\end{align}
where $c^{\dagger}_{i\sigma}$ and $c_{i \sigma}$ are creation and annihilation operators of an itinerant electron at site $i$ and spin $\sigma$, respectively, while $\bm{S}_i$ is a localized spin at site $i$. 
Here, we regard $\bm{S}_i$ as the classical spin with the magnitude of $|\bm{S}_i|=1$. 
The Hamiltonian consists of the kinetic energy term of itinerant electrons in the first term in Eq.~(\ref{eq:Ham_KLM}) and the exchange coupling term between itinerant electron spins $\bm{s}_i$ and localized spins $\bm{S}_i$ in the second term; $\bm{s}_i=(1/2)\sum_{\sigma,\sigma'}c^{\dagger}_{i\sigma} \bm{\sigma}_{\sigma \sigma'} c_{i \sigma'}$ where $\bm{\sigma}=(\sigma^x,\sigma^y,\sigma^z)$ is the vector of Pauli matrices. 
We here do not consider the effect of the spin-orbit coupling by targeting the materials with negligible small spin-orbit coupling, although the extension incorporating such an effect is straightforward. 
$t_{ij}$ and $J>0$ are the hopping and exchange interaction parameters, respectively. 
It is noted that the sign of $J$ is irrelevant in the following result. 

For later convenience, we present the Fourier transform of the model in Eq.~(\ref{eq:Ham_KLM}) as 
\begin{align}
\label{eq:Ham_kspace}
\mathcal{H}=\sum_{\bm{k},\sigma} \varepsilon_{\bm{k}} c^{\dagger}_{\bm{k}\sigma}c_{\bm{k}\sigma} +
\frac{J}{\sqrt{N}} \sum_{\bm{k},\bm{q},\sigma, \sigma'} c^{\dagger}_{\bm{k}\sigma}\bm{\sigma}_{\sigma \sigma'} c_{\bm{k}+\bm{q}\sigma'} \cdot \bm{S}_{\bm{q}}, 
\end{align}
where $c_{\bm{k} \sigma}^{\dagger}$, $c_{\bm{k}\sigma}$, and $\bm{S}_{\bm{q}}$ are the Fourier transform of $c_{i\sigma}^{\dagger}$, $c_{i\sigma}$, and $\bm{S}_i$, respectively, where $N$ is the number of sites. 
$\varepsilon_{\bm{k}}$ is the energy dispersion of the electrons. 
Hereafter, we implicitly consider the centrosymmetric lattice structure without the sublattice degree of freedom for simplicity: $\varepsilon_{\bm{k}}=\varepsilon_{-\bm{k}}$.

\section{Effective momentum-dependent chiral-type spin cross products}
\label{sec:Effective chiral interactions}

From the general symmetry aspect, the necessary conditions of the antisymmetric spin-split band structure are the breakings of spatial inversion symmetry and the product symmetry consisting of spatial inversion and time-reversal symmetries. 
These conditions are naturally satisfied in noncentrosymmetric nonmagnetic systems with the relativistic spin-orbit coupling, such as the Rashba metals. 
Meanwhile, in the absence of the spin-orbit coupling, the spin polarization occurs only when time-reversal symmetry is broken. 
In other words, the spin-dependent band modulation is caused by the scattering due to the ordered localized spins with wave vector $\bm{q}$, i.e., $\langle \bm{S}_{\bm{q}} \rangle \neq 0 $ ($\langle \cdots \rangle$ represents the expectation value), via the exchange coupling in Eq.~(\ref{eq:Ham_kspace}).
Furthermore, the noncollinear spin configurations are necessary to induce the antisymmetric momentum-dependent spin polarization, since the collinear ones do not break spin rotational symmetry, which ensures the twofold degeneracy with respect to the spin degree of freedom~\cite{hayami2019momentum}.

In this section, we examine how the band structures are modulated under noncollinear magnetic orderings with finite ordering wave vectors within the perturbation calculation by supposing that the exchange coupling $J$ is small enough compared to the bandwidth of itinerant electrons. 
We present the momentum-dependent chiral-type bilinear spin cross product in Sec.~\ref{sec:Chiral bilinear interaction} and chiral-type biquadratic spin cross product in Sec.~\ref{sec:Chiral biquadratic interaction}, which are obtained from the second-order and fourth-order contributions in terms of $J$. 
The following results in this section can be applied to any magnetic structures in any lattice systems in one to three spatial dimensions. 
We discuss the result for the specific magnetic textures and the lattices in Sec.~\ref{sec:Antisymmetric spin splitting in single-$Q$ and multiple-$Q$ states}. 

\begin{figure}[t!]
\begin{center}
\includegraphics[width=1.0 \hsize]{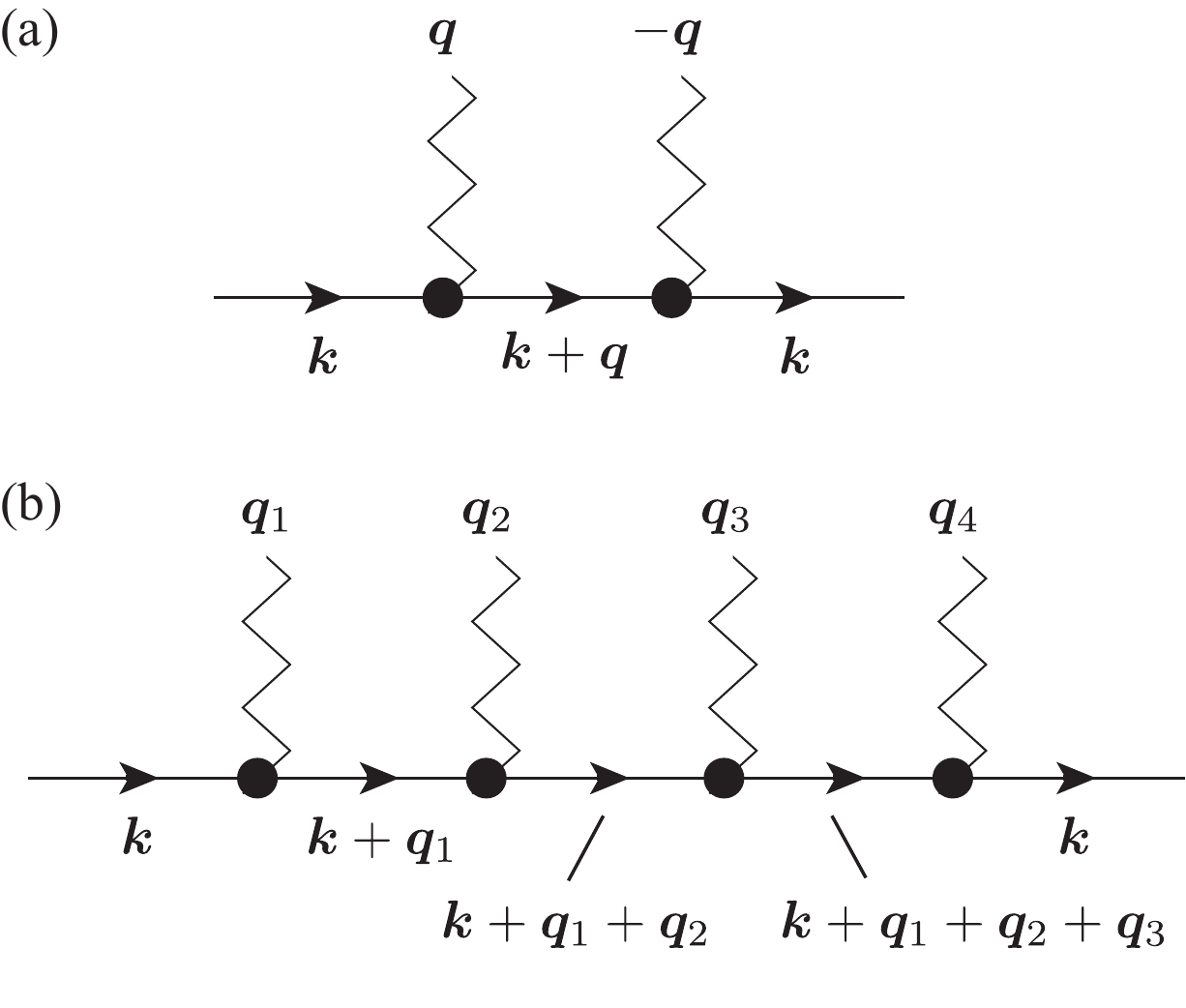} 
\caption{
\label{Fig:diagram}
Feynman diagrams for (a) the second-order and (b) four-order contributions to the $\bm{k}$-resolved spin $\bm{s}_{\bm{k}}$ in the perturbation expansion in terms of the spin-charge coupling $J$. 
See Eqs.~(\ref{eq:perturbation_sq}) and (\ref{eq:perturbation_sq2}) for specific expressions. 
The vertices with wavy lines denote the scattering of the itinerant electrons by the localized spins in the momentum-space representation, and the solid lines with arrows represent the bare propagators of itinerant electrons.
}
\end{center}
\end{figure}

\subsection{Bilinear spin cross product}
\label{sec:Chiral bilinear interaction}

To examine the antisymmetric spin-dependent modulation in the electronic band structure under noncollinear magnetic orderings with finite ordering wave vectors, we evaluate the expectation value of the itinerant electron spin operator $\bm{s}_{\bm{k}}=(1/2)\sum_{\sigma,\sigma'}c^{\dagger}_{\bm{k}\sigma} \bm{\sigma}_{\sigma \sigma'} c_{\bm{k} \sigma'}$ with wave vector $\bm{k}$. 
The lowest-order contribution with respect to $J$ is of second-order, which is derived as 
\begin{align}
\label{eq:perturbation_sq}
\bm{s}_{\bm{k}}=&i\frac{J^2}{N}T\sum_{\bm{q}}\sum_{\omega_n}G^2_{\bm{k}}G_{\bm{k}+\bm{q}} (\bm{S}_{\bm{q}}\times \bm{S}_{-\bm{q}}), 
\end{align}
where $G_{\bm{k}}(i \omega_n)=[i \omega_n-(\varepsilon_{\bm{k}}-\mu)]^{-1}$ is the noninteracting Green's function with the Matsubara frequency $\omega_n$  and $\mu$ is the chemical potential. 
The spin dependence of the Green's function is omitted owing to the absence of the spin-orbit coupling (or spin-dependent hopping) in the model in Eq.~(\ref{eq:Ham_kspace}). 
The summation of the Matsubara frequency can be taken analytically [see Eq.~(\ref{eq:omegasum})].
The corresponding Feynman diagram is shown in Fig.~\ref{Fig:diagram}(a).

The expression in Eq.~(\ref{eq:perturbation_sq}) indicates that the spin cross product, which we call the chiral-type bilinear spin cross product, in momentum space is related to the momentum-dependent spin polarization in the magnetic orderings with finite ordering vector $\bm{q}$; the spin polarization is induced along the direction of $\bm{S}_{\bm{q}}\times \bm{S}_{-\bm{q}}$. 
By using the relation as $\varepsilon_{\bm{k}}=\varepsilon_{-\bm{k}}$, one finds that $\bm{s}_{-\bm{k}}=-\bm{s}_{\bm{k}}$ is satisfied, which means that the $\bm{k}$-antisymmetric spin polarization appears for nonzero $\bm{S}_{\bm{q}}\times \bm{S}_{-\bm{q}}$ and there is no uniform component $\sum_{\bm{k}}\bm{s}_{\bm{k}}=\bm{0}$. 
In other words, a nonzero antisymmetric contribution to $\bm{s}_{\bm{k}}$ appears when the magnetic orderings with nonzero $\bm{S}_{\bm{q}}\times \bm{S}_{-\bm{q}}$ occur.
The momentum dependence of the spin polarization is determined by the product of the Green's function $G^2_{\bm{k}}G_{\bm{k}+\bm{q}}$.

The result in Eq.~(\ref{eq:perturbation_sq}) is reasonable from the symmetry aspect, since the spin cross product $\bm{S}_{\bm{q}}\times \bm{S}_{-\bm{q}}$ becomes nonzero only when spatial inversion symmetry is absent in the system.
For example, the form of $\bm{S}_{\bm{q}}\times \bm{S}_{-\bm{q}}$ appears in the model Hamiltonian as the interaction for noncentrosymmetric itinerant magnets with the spin-orbit coupling~\cite{Hayami_PhysRevLett.121.137202,Okumura_PhysRevB.101.144416,hayami2021field}. 
In contrast, the present bilinear spin cross product is induced by the magnetic orderings which simultaneously break spatial inversion symmetry and does not require the spin-orbit coupling. 
Thus, the antisymmetric spin polarization in this mechanism only appears in the presence of the magnetic orderings; the antisymmetric band structure emerges below the critical temperature in the materials.

From the expression of bilinear spin cross product in momentum space, one finds that nonzero $\bm{S}_{\bm{q}}\times \bm{S}_{-\bm{q}}$ can be obtained by noncollinear and/or noncoplanar spin textures not the collinear ones. 
Moreover, $\bm{S}_{\bm{q}}$ must have both the real and imaginary components for nonzero $\bm{S}_{\bm{q}}\times \bm{S}_{-\bm{q}}$ owing to $\bm{S}_{\bm{q}}=\bm{S}^*_{-\bm{q}}$. 
A simple spin texture to satisfy these conditions is a single-$Q$ spiral one characterized by $\bm{S}_i=(\sin \bm{Q}\cdot \bm{r}_i, \cos \bm{Q}\cdot \bm{r}_i, 0)$ with the position vector $\bm{r}_i$, which induces nonzero $s^z_{\bm{k}} \propto (\bm{S}_{\bm{Q}}\times \bm{S}_{-\bm{Q}})^z$, as will be discussed in Sec.~\ref{sec:Single-$Q$ cycloidal state}. 
We also show that the antisymmetric spin-split band structure is caused by the emergence of the multiple-$Q$ spiral orderings, such as the meron-antimeron and skyrmion crystals, as discussed in Sec.~\ref{sec:Double-$Q$ states}.

\subsection{Biquadratic spin cross product}
\label{sec:Chiral biquadratic interaction}

Similarly, the fourth-order contribution to $\bm{s}_{\bm{k}}$ with respect to $J$ is given by 
\begin{align}
\label{eq:perturbation_sq2}
\bm{s}_{\bm{k}}=&i\frac{J^4}{N^2}T\sum_{\bm{q}_1,\bm{q}_2,\bm{q}_3,\bm{q}_4}\sum_{\omega_n,l}  G^2_{\bm{k}}G_{\bm{k}+\bm{q}_1}G_{\bm{k}+\bm{q}_1+\bm{q}_2}G_{\bm{k}+\bm{q}_1+\bm{q}_2+\bm{q}_3} \nonumber \\
&\times \delta_{\bm{q}_1+\bm{q}_2+\bm{q}_3+\bm{q}_4,l \bm{G}}
\big[(
\bm{S}_{\bm{q}_1}\times \bm{S}_{\bm{q}_2})(\bm{S}_{\bm{q}_3}\cdot \bm{S}_{\bm{q}_4}) \nonumber \\
&+(\bm{S}_{\bm{q}_3}\times \bm{S}_{\bm{q}_4})(\bm{S}_{\bm{q}_1}\cdot \bm{S}_{\bm{q}_2})
+(\bm{S}_{\bm{q}_1}\times \bm{S}_{\bm{q}_4})(\bm{S}_{\bm{q}_2}\cdot \bm{S}_{\bm{q}_3})  \nonumber \\
&+(\bm{S}_{\bm{q}_2}\times \bm{S}_{\bm{q}_3})(\bm{S}_{\bm{q}_1}\cdot \bm{S}_{\bm{q}_4})
-(\bm{S}_{\bm{q}_1}\times \bm{S}_{\bm{q}_3})(\bm{S}_{\bm{q}_2}\cdot \bm{S}_{\bm{q}_4})  \nonumber \\
&-(\bm{S}_{\bm{q}_2}\times \bm{S}_{\bm{q}_4})(\bm{S}_{\bm{q}_1}\cdot \bm{S}_{\bm{q}_3})
\big], 
\end{align}
where $\delta$ is the Kronecker delta and $\bm{G}$ is the reciprocal lattice vector ($l$ is an integer).
In this case, the summation in terms of the Matsubara frequency is calculated for a certain temperature, and then, it is taken in the $T \to 0$ limit.
The corresponding Feynman diagram is shown in Fig.~\ref{Fig:diagram}(b).

Equation~(\ref{eq:perturbation_sq2}) gives the four-spin cross product. 
The functional form of the four-spin cross product resembles the chiral biquadratic interaction in real space described by $(\bm{S}_i \times \bm{S}_j)(\bm{S}_i\cdot \bm{S}_j)$~\cite{brinker2019chiral,Laszloffy_PhysRevB.99.184430,Mankovsky_PhysRevB.101.174401,Brinker_PhysRevResearch.2.033240,lounis2020multiple,Dos_PhysRevB.103.L140408}. 
In the present case, however, the biquadratic spin cross product is defined for the Fourier components of spins and becomes nonzero only under the magnetic orderings. 
In other words, the present biquadratic spin cross product only contributes to the momentum-dependent polarization and does not contribute to the free energy.
Similar to the bilinear spin cross product, $\bm{s}_{\bm{k}}$ from the biquadratic spin cross product satisfy $\sum_{\bm{k}}\bm{s}_{\bm{k}}=\bm{0}$.
The four-spin cross product in Eq.~(\ref{eq:perturbation_sq2}) can account for the antisymmetric spin-polarization in the multiple-$Q$ states that are not explained by the bilinear spin cross product in Eq.~(\ref{eq:perturbation_sq}), as shown in Sec.~\ref{sec:Other double-$Q$ state}.

\section{Antisymmetric spin splitting in single-$Q$ and multiple-$Q$ states}
\label{sec:Antisymmetric spin splitting in single-$Q$ and multiple-$Q$ states}

We discuss the antisymmetric spin splitting in the band structure in the presence of magnetic orderings. 
As the result in Sec.~\ref{sec:Effective chiral interactions} can be applied to any lattice structures in one to three spatial dimensions, we here show the examples in the one- and two-dimensional cases.
First, we show the results in the single-$Q$ cycloidal spiral state in the one-dimensional chain in Sec.~\ref{sec:Single-$Q$ cycloidal state}. 
Then, we discuss the antisymmetric spin splittings in the two double-$Q$ states, the meron-antimeron and skyrmion crystals in Sec.~\ref{sec:Double-$Q$ states}, and in the other double-$Q$ noncoplanar state in Sec.~\ref{sec:Other double-$Q$ state} on the two-dimensional square lattice. 
Although we here discuss the magnetic orderings only with the commensurate ordering vectors, a qualitative similar result is obtained when the modulation vectors are incommensurate, as clearly found in Eqs.~(\ref{eq:perturbation_sq}) and (\ref{eq:perturbation_sq2}).
As shown in each example, the antisymmetric spin polarizations are well explained by the expressions in Eqs.~(\ref{eq:perturbation_sq}) and (\ref{eq:perturbation_sq2}). 

\subsection{Single-$Q$ cycloidal spiral state}
\label{sec:Single-$Q$ cycloidal state}

\begin{figure}[htb!]
\begin{center}
\includegraphics[width=1.0 \hsize]{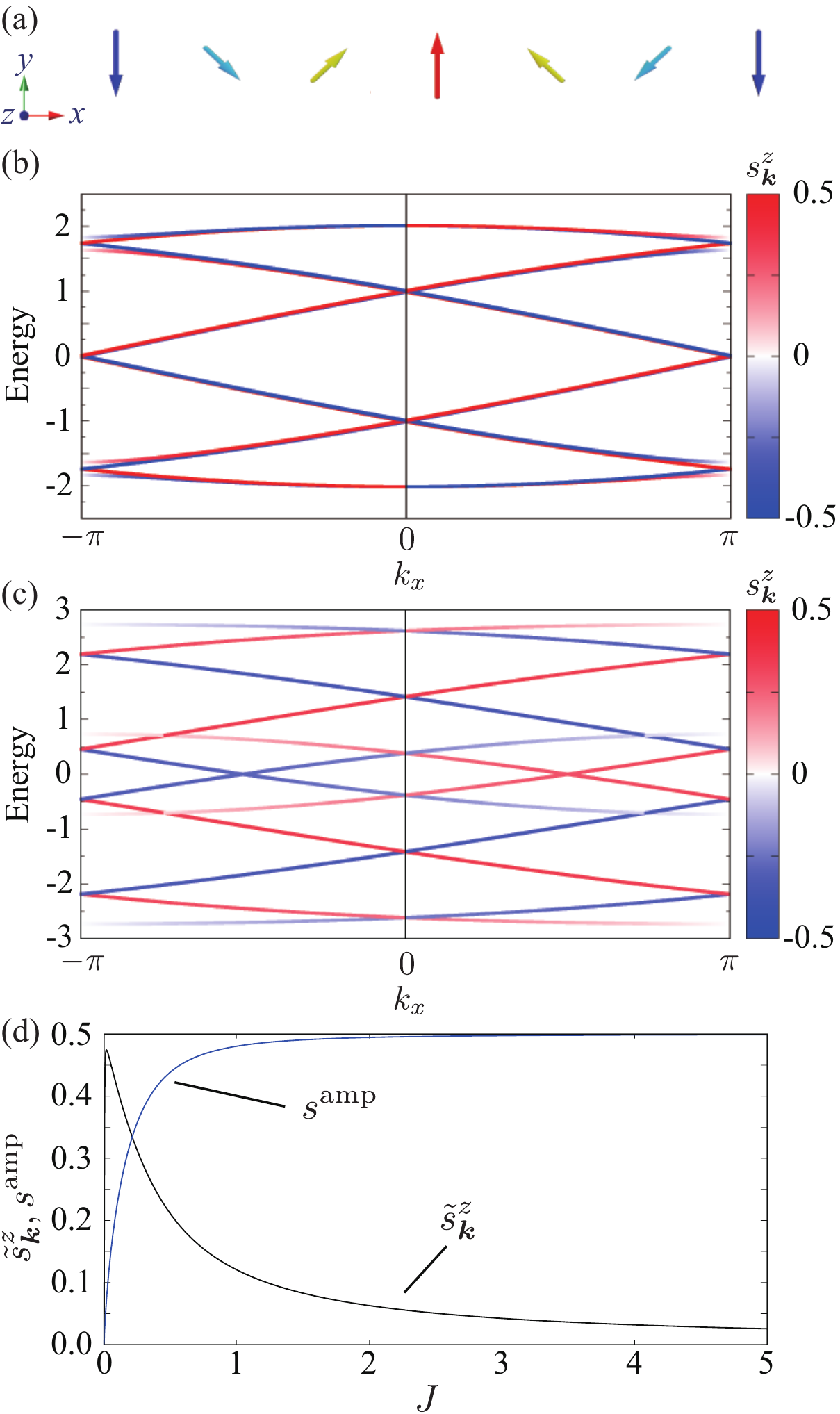} 
\caption{
\label{Fig:band_1D}
(a) Cycloidal spiral state. 
(b), (c) The band structures at (b) $J=0.1$ and (c) $J=1$. 
The color map shows the spin polarization of the $z$ component at each wave vector. 
(d) $J$ dependence of $\tilde{s}^z_{\bm{k}}=(2/N)\sum_{k_x>0} s^z_{\bm{k}}$ and $s^{\rm amp}=\sqrt{\sum_i\left[ (s^x_i)^2+(s^y_i)^2\right]}$ for the lowest band. 
}
\end{center}
\end{figure}

We consider the single-$Q$ cycloidal spiral state on the one-dimensional chain along the $x$ direction, where we take the lattice constant as unity. 
The spin configuration is given by $\bm{S}_i=(\sin Q x_i, \cos Q x_i, 0)$ with $Q=\pi/3$, whose schematic picture is shown in Fig.~\ref{Fig:band_1D}(a).  
In the following, we take the nearest-neighbor hopping $t_1=1$ in $\varepsilon_{\bm{k}}$. 

Figure~\ref{Fig:band_1D}(b) shows the band structure at $J=0.1$. 
The color map shows the spin polarization of the $z$ component, where the red (blue) lines show the positive (negative) $z$-spin component; momentum dependence of the spin splitting is represented by $s^z_{\bm{k}}\propto k_x \sigma^z$. 
Although the antisymmetric spin polarization in the band structure is similar to that in the noncentrosymmetric system with the Rashba spin-orbit coupling, but the origin of the antisymmetric spin polarization is different with each other.
The present antisymmetric spin polarization is caused by the single-$Q$ cycloidal spiral ordering without the spin-orbit coupling. 
Such a behavior remains for large $J$, as shown in Fig.~\ref{Fig:band_1D}(c) in the case of $J=1$. 

The microscopic origin of the antisymmetric spin polarization is understood from the bilinear spin cross product in Eq.~(\ref{eq:perturbation_sq}). 
From the spiral spin configuration, we find the $z$ component of the antisymmetric spin polarization as $s^z_{\bm{k}} \propto (\bm{S}_{Q}\times \bm{S}_{-Q})^z$. 
Besides, the momentum dependence of the spin polarization is given by the factor $G^2_{k_x}G_{k_x+Q}$. 
When using the following relation by eliminating the summation with respect to the Matsubara frequency as 
\begin{align}
\label{eq:omegasum}
T\sum_{\omega_n}G^2_{k_x}G_{k_x+Q}&= \frac{f(\varepsilon_{k_x+Q})-f(\varepsilon_{k_x})}{(\varepsilon_{k_x}-\varepsilon_{k_x+Q})^2} \nonumber \\
&\ \ \ \ +\frac{1}{\varepsilon_{k_x}-\varepsilon_{k_x+Q}}\frac{{\rm d}f(\varepsilon_{k_x})}{{\rm d}\varepsilon_{k_x}}, \\
\varepsilon_{k_x}&=-2t_1\cos k_x , 
\end{align}
we can evaluate the $k_x$ dependence of the antisymmetric spin polarization. 

The degree of the antisymmetric spin polarization depends on the amplitude of the order parameters and the band structure. 
To demonstrate that, we show the $J$ dependence of $\tilde{s}^z_{\bm{k}}=(2/N)\sum_{k_x>0} s^z_{\bm{k}}$ for the lowest band in Fig.~\ref{Fig:band_1D}(d), where the small (large) $J$ regime mimics the situation with the small (large) order parameters. 
$\tilde{s}^z_{\bm{k}}$ becomes nonzero for $J>0$ and shows the maxima at $J \simeq 0.018$. 
The increment of $\tilde{s}^z_{\bm{k}}$ for $0<J \lesssim 0.018$ is owing to the enhancement of the spin moment of conduction electrons $s^{\rm amp} = \sqrt{\sum_i\left[ (s^x_i)^2+(s^y_i)^2\right]}$. 
Meanwhile, the suppression of $\tilde{s}^z_{\bm{k}}$ for $0.018 \lesssim J$ might be attributed to the electronic band structure where the lowest band tends to be decoupled from the other bands while increasing $J$ [see Figs.~\ref{Fig:band_1D}(a) and \ref{Fig:band_1D}(b)], and hence, the denominator in Eq.~(\ref{eq:omegasum}) becomes large. 

It is noted that a similar antisymmetric spin polarization occurs in the magnetic ordering with the elliptical spiral $\bm{S}_i=(a_x \sin Q x_i, a_y \cos Q x_i, 0)$ where $a_x \neq a_y$. 
Meanwhile, the antisymmetric spin polarization vanishes in the collinear sinusoidal case, i.e., $a_x=0$ or $a_y=0$ due to $\bm{S}_{Q}\times \bm{S}_{-Q}=\bm{0}$

\subsection{Double-$Q$ spiral states}
\label{sec:Double-$Q$ states}

\begin{figure*}[htb!]
\begin{center}
\includegraphics[width=1.0 \hsize]{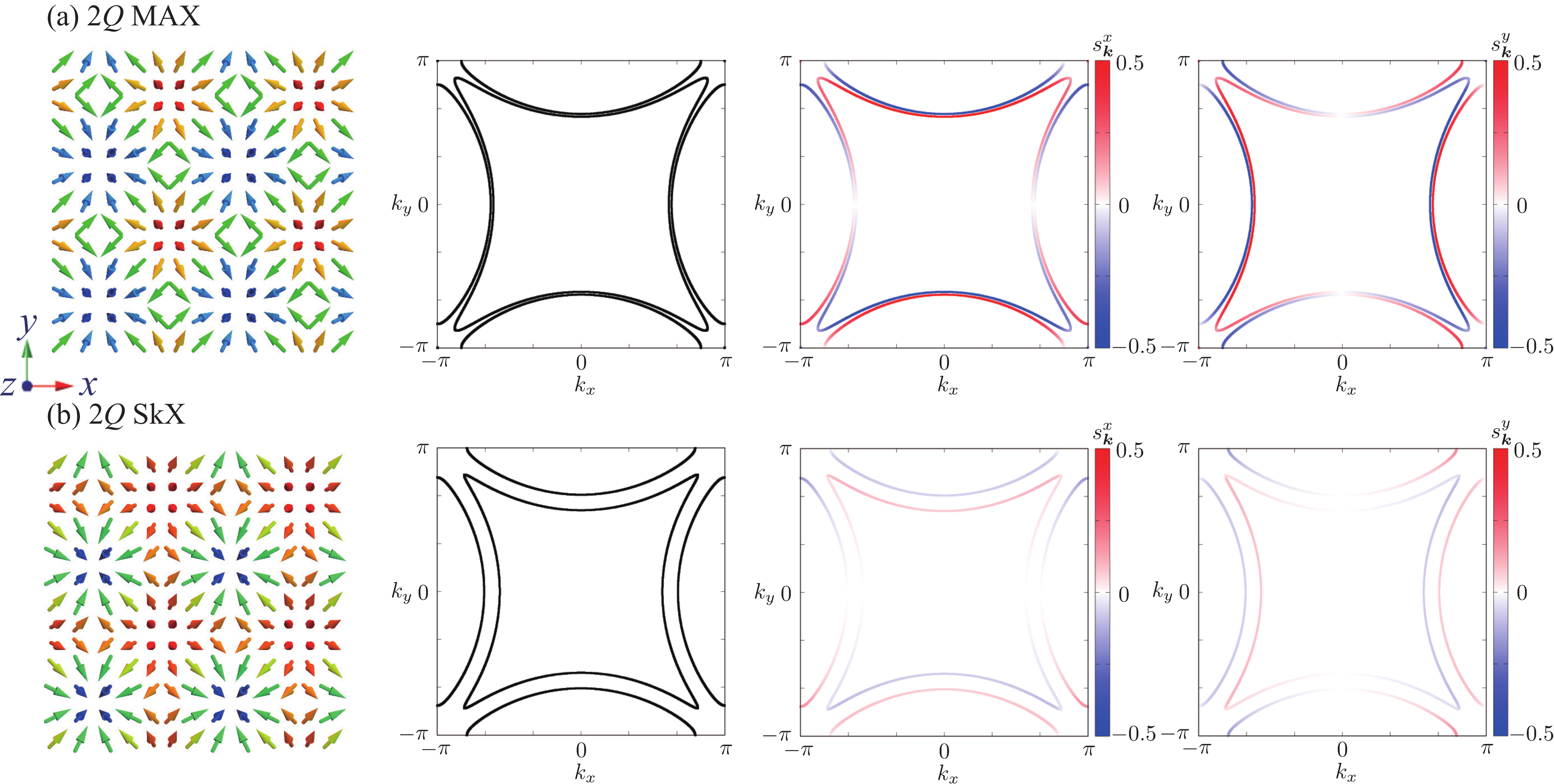} 
\caption{
\label{Fig:band_MAX}
(Left panel) Real-space spin configurations of (a) the double-$Q$ meron-antimeron crystal (2$Q$ MAX) in Eq.~(\ref{eq:2QMAX}) and (b) the double-$Q$ skyrmion crystal (2$Q$ SkX) in Eq.~(\ref{eq:2QSkX}). 
(Middle left panel) The isoenergy surfaces at $J=0.1$ and $\mu=3.5$ in the Brillouin zone. 
(Middle right and right panels) The spin polarization of the $x$ and $y$ components at each wave vector corresponding to the middle left panel. 
}
\end{center}
\end{figure*}

The above analysis can be directly applied to multiple-$Q$ states, which consists of multiple  spiral waves. 
We consider two double-$Q$ states described by superposing of single-$Q$ cycloidal spirals, the meron-antimeron crystal in Sec.~\ref{sec:Meron-antimeron crystal} and the skyrmion crystal in Sec.~\ref{sec:Skyrmion crystal}, on the square lattice with the nearest-neighbor hopping $t_1=1$. 
We take the ordering vectors $\bm{Q}_1=(\pi/3,0)$ and $\bm{Q}_2=(0,\pi/3)$, which are connected by fourfold rotational symmetry. 

\subsubsection{Meron-antimeron crystal}
\label{sec:Meron-antimeron crystal}

The meron-antimeron crystal is represented by a superposition of two cycloidal spirals. 
The real-space spin configuration is given by 
\begin{align}
\label{eq:2QMAX}
\bm{\tilde{S}}_i &=
\left(
    \begin{array}{c}
     \cos  \bm{Q}_1 \cdot \bm{r}_i  \\
    \cos  \bm{Q}_2 \cdot \bm{r}_i  \\
 -\sin  \bm{Q}_1 \cdot \bm{r}_i  -\sin  \bm{Q}_2 \cdot \bm{r}_i 
          \end{array}
  \right)^{\rm T}, \nonumber \\
  \bm{S}_i&=\frac{\bm{\tilde{S}}_i}{|\bm{\tilde{S}}_i|}. 
\end{align}
The schematic spin configuration is shown in the left panel of Fig.~\ref{Fig:band_MAX}(a). 
Upon close looking in real space, one finds that the spin configuration consists of a periodic array of meron and antimeron with an opposite-sign half skyrmion number~\cite{brey1996skyrme,yu2018transformation,kurumaji2019skyrmion}. 
The stabilization mechanism of the meron-antimeron crystal has been widely studied in chiral insulating magnets~\cite{Lin_PhysRevB.91.224407}, frustrated insulating  magnets~\cite{Wang_PhysRevB.103.104408}, and polar itinerant magnets~\cite{Hayami_PhysRevLett.121.137202,hayami2021meron}. 

As the spin configuration in Eq.~(\ref{eq:2QMAX}) is characterized by two spiral waves along the different directions, the antisymmetric spin polarization occurs for the two spin components on the basis of the bilinear spin cross product in Eq.~(\ref{eq:perturbation_sq}): 
One is the $y$-spin antisymmetric polarization that arises from the $\bm{Q}_1$ spiral [$(\bm{S}_{\bm{Q}_1}\times \bm{S}_{-\bm{Q}_1})^y \neq 0$] and the other is the $x$-spin one that arises from the $\bm{Q}_2$ spiral [$(\bm{S}_{\bm{Q}_2}\times \bm{S}_{-\bm{Q}_2})^x \neq 0$]. 
Then, one finds that the total antisymmetric spin polarization in the band structure is described by $-k_x \sigma_y +k_y \sigma_x$. 
Indeed, the functional form of the antisymmetric spin polarization obtained by the direct diagonalization is consistent with that by Eq.~(\ref{eq:perturbation_sq}), as shown in the right two figures in Fig.~\ref{Fig:band_MAX}(a), where we also plot the isoenergy surfaces at $\mu=3.5$ in the Brillouin zone in the middle left panel for reference. 

From the symmetry viewpoint, the functional form of $-k_x \sigma_y +k_y \sigma_x$ is the same as that induced by the Rashba-type spin-orbit coupling under the polar point group where the electric dipole moment is activated. 
This is reasonable, since the real-space magnetic texture has the same symmetry as the electric dipole moment along the $z$ direction; spatial inversion symmetry and mirror symmetry in terms of the horizontal plane are broken. 
Thus, this mechanism to induce the antisymmetric spin polarization is regarded as the inverse antisymmetric spin polarization mechanism, which is analogous to the inverse Dzyaloshinskii-Moriya mechanism~\cite{tokura2014multiferroics}.

\subsubsection{Skyrmion crystal}
\label{sec:Skyrmion crystal}

We show the antisymmetric spin polarization in the double-$Q$ skyrmion crystal, whose spin configuration is given by 
\begin{align}
\label{eq:2QSkX}
\bm{\tilde{S}}_i &=
\left(
    \begin{array}{c}
     \cos  \bm{Q}_1 \cdot \bm{r}_i  \\
    \cos  \bm{Q}_2 \cdot \bm{r}_i  \\
\tilde{M}^z   -\sin  \bm{Q}_1 \cdot \bm{r}_i  -\sin  \bm{Q}_2 \cdot \bm{r}_i 
          \end{array}
  \right)^{\rm T}, \nonumber \\
    \bm{S}_i&=\frac{\bm{\tilde{S}}_i}{|\bm{\tilde{S}}_i|}, 
\end{align}
where $\tilde{M}^z=0.7$. 
In contrast to the spin texture of the meron-antimeron crystal in Fig.~\ref{Fig:band_MAX}(a), the region with the positive (negative) $S^z_i$ extends (shrinks) owing to the introduction of $\tilde{M}^z$, as shown in the left panel of Fig.~\ref{Fig:band_MAX}(b).  
As a result, the skyrmion crystal exhibits the topological Hall effect. 
The 2$Q$ skyrmion crystal appears in the ground state in itinerant magnets~\cite{Hayami_PhysRevB.103.024439,Hayami_doi:10.7566/JPSJ.89.103702} and in localized magnets~\cite{Utesov_PhysRevB.103.064414,Wang_PhysRevB.103.104408}. 

The right three panels of Fig.~\ref{Fig:band_MAX}(b) shows the isoenergy surfaces at $J=0.1$ and $\mu=3.5$, where the right two panels show the spin polarization of the $x$- and $y$-spin components at each wave vector. 
The behavior of the antisymmetric spin polarization is similar to that in the meron-antimeron crystal in Fig.~\ref{Fig:band_MAX}(a). 
This is because the difference between the meron-antimeron crystal and the skyrmion crystal is in the nonzero uniform $z$-spin component while keeping the double-$Q$ spiral spin texture, which does not lead to a qualitative difference. 
The same discussion holds when considering the double-$Q$ spin texture with large $\tilde{M}^z$ so that the spin texture has no skyrmion number; the same antisymmetric spin polarization in the form of $-k_x \sigma_y +k_y \sigma_x$ occurs unless $\tilde{S}^x_i=\tilde{S}^y_i=0$ in Eq.~(\ref{eq:2QSkX}).

\subsection{Other double-$Q$ state}
\label{sec:Other double-$Q$ state}

\begin{figure}[htb!]
\begin{center}
\includegraphics[width=1.0 \hsize]{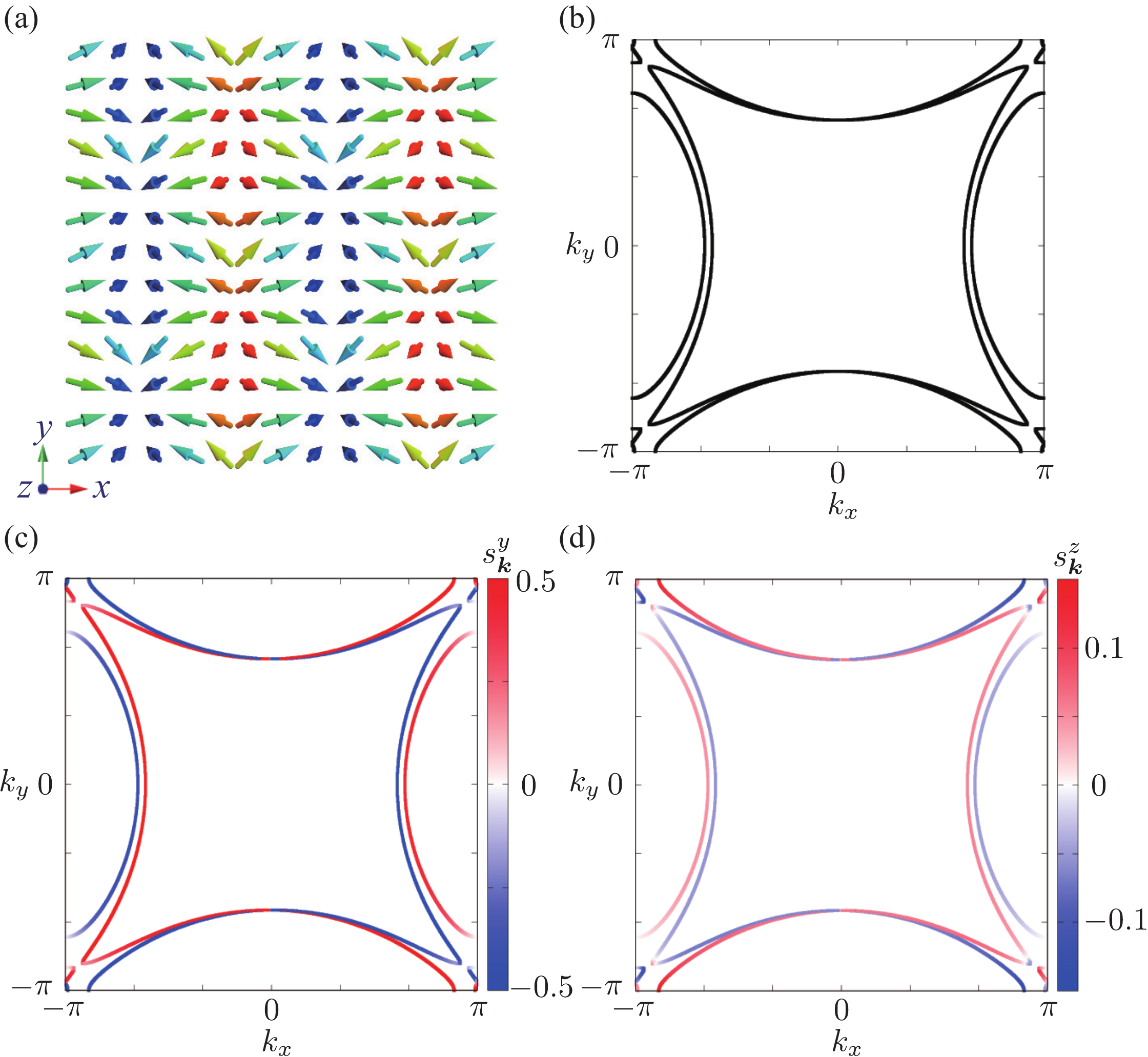} 
\caption{
\label{Fig:band_other}
(a) Real-space spin configurations of the $2Q$ noncoplanar state at $b=0.5$ in Eq.~(\ref{eq:2Qnoncop}). 
(b) The isoenergy surfaces at $J=0.1$ and $\mu=3.5$ in the Brillouin zone. 
(c), (d) The spin polarization of the (c) $y$ and (d) $z$ components at each wave vector. 
}
\end{center}
\end{figure}

In the previous sections (Secs.~\ref{sec:Single-$Q$ cycloidal state} and \ref{sec:Double-$Q$ states}), we show that the effective bilinear spin cross product under the (multiple-$Q$) spiral orderings give rise to the antisymmetric spin polarization in the band structure.  
In this section, we present the situation where the magnetic orderings exhibit the antisymmetric spin polarization in the presence of the effective biquadratic spin cross product rather than the bilinear one. 
For that purpose, we consider the following double-$Q$ spin configuration as 
\begin{align}
\label{eq:2Qnoncop}
\bm{\tilde{S}}_i &=
\left(
    \begin{array}{c}
     \cos  \bm{Q}_1 \cdot \bm{r}_i  \\
    b \cos  \bm{Q}_2 \cdot \bm{r}_i  \\
-\sin  \bm{Q}_1 \cdot \bm{r}_i  -b \cos  \bm{Q}_2 \cdot \bm{r}_i 
          \end{array}
  \right)^{\rm T},\nonumber \\
    \bm{S}_i&=\frac{\bm{\tilde{S}}_i}{|\bm{\tilde{S}}_i|}, 
\end{align}
where $b$ is the variational parameter to represent the relative amplitude of the second $\bm{Q}_2$ component. 
The spin configuration in this state consists of the spiral along the $\bm{Q}_1$ direction and the sinusoidal modulation along the $\bm{Q}_2$ direction with the different intensities. 
The expression in Eq.~(\ref{eq:2Qnoncop}) becomes equivalent with that of the meron-antimeron crystal in Eq.~(\ref{eq:2QMAX}) when taking $b=1$ and replacing $\cos  \bm{Q}_2 \cdot \bm{r}_i $ in the $z$-spin component with $\sin  \bm{Q}_2 \cdot \bm{r}_i $. 
Reflecting such a difference of the spin configuration from the meron-antimeron crystal, the real-space spin texture is clearly different, as shown in Fig.~\ref{Fig:band_other}(a). 
Here and hereafter, we take $b=0.5$.

Figure~\ref{Fig:band_other}(b) shows the isoenergy surface in the 2$Q$ noncoplanar state in Eq.~(\ref{eq:2Qnoncop}) at $J=0.1$ and $\mu=3.5$, which is similar to that in the meron-antimeron crystal in Fig.~\ref{Fig:band_MAX}(a) except for the regions around $(k_x,k_y)=(\pi, \pi)$.
As the spin texture includes the spiral along the $\bm{Q}_1$ direction, one finds the antisymmetric spin polarization in terms of the $y$-spin component, as shown in Fig.~\ref{Fig:band_other}(c). 
The origin of this antisymmetric spin splitting is well accounted for by the effective bilinear spin cross product under the $\bm{Q}_1$ spiral, as discussed in Sec.~\ref{sec:Single-$Q$ cycloidal state}. 
Meanwhile, no antisymmetric spin polarization occurs in terms of the $x$-spin component (not shown), since the spin oscillation with the $\bm{Q}_2$ component is described by the collinear(sinusoidal)-type oscillation. 

Notably, we find that the antisymmetric spin polarization in terms of the $z$-spin component, as shown in Fig.~\ref{Fig:band_other}(d). 
The origin of this antisymmetric spin polarization is understood by the effective biquadratic spin cross product instead of the bilinear one, since the quantity $(\bm{S}_{\bm{q}_1}\times \bm{S}_{\bm{q}_2})^z(\bm{S}_{\bm{q}_3}\cdot \bm{S}_{\bm{q}_4}) $ in Eq.~(\ref{eq:perturbation_sq2}) becomes nonzero for $(\bm{q}_1,\bm{q}_2,\bm{q}_3,\bm{q}_4)=(\bm{Q}_1, \bm{Q}_2, -\bm{Q}_1, -\bm{Q}_2)$ with the $\bm{k}$-dependent form factor $G^2_{\bm{k}}G_{\bm{k}+\bm{Q}_1}G_{\bm{k}+\bm{Q}_2}G_{\bm{k}+\bm{Q}_1+\bm{Q}_2}$.
The above result clearly indicates that the resultant antisymmetric spin polarization strongly depends on the way of a superposition of the multiple-$Q$ spin density waves. 
In other words, detecting the antisymmetric spin polarization in experiments, such as spin- and angle-resolved photoemission spectroscopy, might be useful to deduce the constituent waves in the multiple-$Q$ states. 

\section{Discussion}
\label{sec:Discussion}
In this section, we discuss the candidate centrosymmetric multiple-$Q$ magnetic materials to exhibit the antisymmetric spin splitting in Sec.~\ref{sec:Relevant materials} and present the expected physical phenomena driven by the momentum-dependent antisymmetric spin polarization in Sec.~\ref{sec:Relevant physical phenomena}. 

\subsection{Relevant materials}
\label{sec:Relevant materials}

The expressions in Eqs.~(\ref{eq:perturbation_sq}) and (\ref{eq:perturbation_sq2}) can be straightforwardly applied to complex noncollinear and noncoplanar spin configurations. 
As an example, we apply the derived expressions to the multiple-$Q$ states observed in the $f$-electron compound GdRu$_2$Si$_2$~\cite{khanh2020nanometric,Yasui2020imaging}. 
The crystal structure of this compound is the centrosymmetric tetragonal crystal structure. 
The Lorentz transmission electron microscopy and spectroscopic-imaging scanning tunneling microscopy measurements have clarified three double-$Q$ states: the double-$Q$ spiral state, the double-$Q$ skyrmion crystal, and the double-$Q$ fan state from the low magnetic-field region. 
The theoretical model calculations have indicated the real-space spin configurations for three double-$Q$ states are given by 
\begin{align}
\label{eq:2Qscrew}
\bm{\tilde{S}}_i &=
\left(
    \begin{array}{c}
   - b \cos  \bm{Q}_2 \cdot \bm{r}_i  \\
     \cos  \bm{Q}_1 \cdot \bm{r}_i  \\
a_z \sin  \bm{Q}_1 \cdot \bm{r}_i  
          \end{array}
  \right)^{\rm T},\nonumber \\
    \bm{S}_i&=\frac{\bm{\tilde{S}}_i}{|\bm{\tilde{S}}_i|}, 
\end{align}
for the double-$Q$ spiral state, 
\begin{align}
\label{eq:2QSkX}
\bm{\tilde{S}}_i &=
\left(
    \begin{array}{c}
    - \cos  \bm{Q}_2 \cdot \bm{r}_i  \\
    \cos  \bm{Q}_1 \cdot \bm{r}_i  \\
\tilde{M}^z   -\sin  \bm{Q}_1 \cdot \bm{r}_i  -\sin  \bm{Q}_2 \cdot \bm{r}_i 
          \end{array}
  \right)^{\rm T}, \nonumber \\
    \bm{S}_i&=\frac{\bm{\tilde{S}}_i}{|\bm{\tilde{S}}_i|}, 
\end{align}
for the double-$Q$ skyrmion crystal, and
\begin{align}
\label{eq:2Qfan}
\bm{\tilde{S}}_i &=
\left(
    \begin{array}{c}
      - \cos  \bm{Q}_2 \cdot \bm{r}_i  \\
    \cos  \bm{Q}_1 \cdot \bm{r}_i   \\
\tilde{M}^z    
          \end{array}
  \right)^{\rm T}, \nonumber \\
    \bm{S}_i&=\frac{\bm{\tilde{S}}_i}{|\bm{\tilde{S}}_i|}, 
\end{align}
for the double-$Q$ fan state~\cite{Hayami_PhysRevB.103.024439}. 
$b$, $a_z$, and $\tilde{M}^z$ are appropriate constants depending on the magnetic field. 
The spin configuration of the double-$Q$ skyrmion crystal is similar to that in Eq.~(\ref{eq:2QSkX}); the difference is found in the helicity.

Although all the three states are characterized by the double-$Q$ spin configurations, 
the resultant antisymmetric spin polarization is different with each other: 
The double-$Q$ spiral state exhibits the antisymmetric spin polarization in the form of $k_x \sigma_x$, the skyrmion crystal shows the antisymmetric spin polarization in the form of $k_x \sigma_x +k_y \sigma_y$, and the double-$Q$ fan state shows no antisymmetric spin polarization. 
The difference of the antisymmetric spin polarization between three magnetic states is explained by Eq.~(\ref{eq:perturbation_sq}), which was confirmed by the direct diagonalization of the Hamiltonian (not shown). 
Thus, the spin- and angle-resolved photoemission spectroscopy measurement is another experimental probe to distinguish the multiple-$Q$ spin textures including the helicity of the skyrmion crystal in addition to the Lorentz transmission electron microscopy and spectroscopic-imaging scanning tunneling microscopy measurements.

The appearance of the antisymmetric spin polarization is also expected in the other centrosymmetric multiple-$Q$ magnets, such as the skyrmion-hosting triangular and kagome magnets~\cite{kurumaji2019skyrmion} and the hedgehog-hosting cubic magnets~\cite{hirschberger2019skyrmion,Hirschberger_10.1088/1367-2630/abdef9}.
In addition, the relation between the spin cross products and the antisymmetric spin polarization can be extended to noncentrosymmetric magnets. 
Although there are two contributions from the spin-orbit coupling and the magnetic orderings to the antisymmetric spin polarization in noncentrosymmetric magnets, the antisymmetric spin polarization by the magnetic orderings occurs only below the transition temperature. 
Thus, the comparison of the spin-split band structures above and below the transition temperature in experiments would provide information about the constituent waves in the multiple-$Q$ states. 
Such information will provide a clue to understand unidentified magnetic orderings in GdSb$_x$Te$_{2-x-\delta}$~\cite{Shiming_PhysRevB.103.134418}, EuAl$_4$~\cite{onuki2020unique,Shang_PhysRevB.103.L020405,kaneko2021charge}, EuGa$_2$Al$_2$~\cite{moya2021incommensurate}, EuGa$_4$~\cite{zhang2021giant}, and EuPtAS~\cite{xie2021complex}. 

\subsection{Relevant physical phenomena}
\label{sec:Relevant physical phenomena}

The antisymmetric spin polarization means an effective coupling between the spin and the momentum in itinerant electrons, which is called the spin-momentum locking~\cite{hsieh2009tunable}. 
Although the spin-momentum locking and its related physical phenomena have been often discussed in noncentrosymmetric nonmagnetic systems with the Rashba and Dresselhaus spin-orbit interaction as discussed in the introduction, similar physical phenomena can be expected in the present magnetic-order-driven antisymmetric spin polarization. 
One of the example is the Edelstein effect where the uniform magnetization $M_\mu$ is induced by applying an electric current $J_\nu$~\cite{edelstein1990spin}, i.e., $M_\mu=\sum_\nu \alpha_{\mu\nu}J_{\nu}$ for $\mu,\nu=x,y,z$.
The tensor $\alpha_{\mu\nu}$ becomes nonzero in the presence of the antisymmetric spin polarization $k_\nu \sigma_{\mu}$. 
Another example is the nonlinear Hall effect on the basis of the Berry curvature dipole mechanism~\cite{Sodemann_PhysRevLett.115.216806,Nandy_PhysRevB.100.195117}. 
In a similar way, other physical phenomena induced by the inversion symmetry breaking are found in the multiple-$Q$ states with nonzero $\bm{S}_{\bm{q}}\times \bm{S}_{-\bm{q}}$ and $(\bm{S}_{\bm{q}_1}\times \bm{S}_{\bm{q}_2})(\bm{S}_{\bm{q}_3}\cdot \bm{S}_{\bm{q}_4})$. 

\section{Summary}
\label{sec:Summary}

To summarize, we have investigated the antisymmetric spin polarization in the band structure induced by the single-$Q$ and multiple-$Q$ spiral orderings. 
By performing the perturbation calculation with respect to the spin-charge coupling in the classical Kondo lattice model, we find that effective chiral-type bilinear and biquadratic spin cross products in momentum space are related to the antisymmetric spin-split band structure in the absence of the relativistic spin-orbit coupling. 
The obtained expressions indicate that the antisymmetric spin polarization occurs in the spin component perpendicular to the spiral plane and the momentum dependence is determined by the product of the Green's function of itinerant electrons. 
We demonstrate the presence of the antisymmetric spin splittings in the single-$Q$ state on the one-dimensional chain and the three double-$Q$ states, the meron-antimeron crystal, the skyrmion crystal, and the noncoplanar state, on the two-dimensional square lattice. 
We show that a way of superposing of the multiple spin density waves leads to a qualitatively different antisymmetric spin polarization. 
We also discuss the relevant materials and physical phenomena under the present antisymmetric spin polarization induced by the magnetic phase transitions. 

The results open up a possibility of engineering the giant antisymmetric spin splitting without relying on the relativistic spin-orbit coupling~\cite{Hayami_PhysRevB.101.220403,Hayami_PhysRevB.102.144441}. 
The obtained expressions in Eqs.~(\ref{eq:perturbation_sq}) and (\ref{eq:perturbation_sq2}) are applied to various itinerant electron systems irrespective of the lattice and magnetic structures. 
As the resultant antisymmetric spin polarization is qualitatively similar to that by the noncentrosymmetric nonmagnetic system with the spin-orbit coupling, the emergence of the Edelstein effect and the nonreciprocal transport is expected, as discussed in Sec.~\ref{sec:Relevant physical phenomena}. 
Thus, the present results will provide a way of bottom-up design approach to realize parity-violating physical phenomena on the basis of the spiral magnetic textures. 

A close relation between the real-space spin texture and momentum-space spin polarization might be provide a deep understanding of the multiple-$Q$ magnetism. 
For example, it is possible to obtain information about a way of superposing the spin density waves by detecting the spin-dependent electronic band structure based on spin- and angle-resolved photoemission spectroscopy. 
Such an indirect identification of the magnetic textures through electric probes like the spectroscopic imaging scanning tunneling microscopy measurement and the noncolinear magnetoresistance has been performed in magnetic materials with complicated magnetic textures~\cite{hanneken2015electrical,Kubetzka_PhysRevB.95.104433,Yasui2020imaging,hayami2021charge}. 
The observation of the momentum-dependent antisymmetric spin polarization would be also useful to understand the nature of the multiple-$Q$ states in both real and momentum spaces.

\begin{acknowledgments}
This research was supported by JSPS KAKENHI Grants Numbers JP19K03752, JP19H01834, JP21H01037, and by JST PRESTO (JPMJPR20L8). 
Parts of the numerical calculations were performed in the supercomputing systems in ISSP, the University of Tokyo.
\end{acknowledgments}

\bibliographystyle{apsrev}
\bibliography{ref.bib}

\end{document}